\documentclass[%
 reprint,
 amsmath,amssymb,
 aps,
prd,
floatfix,
]{revtex4-1}

\usepackage{graphicx}
\usepackage{dcolumn}
\usepackage{bm}
\usepackage{slashed}
\usepackage{color}
\usepackage{textcomp}
\usepackage{enumerate}
\usepackage{extarrows}
\usepackage{comment}
\usepackage{float}

\usepackage{tikz}
\usetikzlibrary{calc} 
\usetikzlibrary{patterns} 
\usetikzlibrary{decorations.pathreplacing} 
\usetikzlibrary{decorations.markings} 
\usetikzlibrary{decorations.pathmorphing} 
\usetikzlibrary{positioning}

\newcommand{\eq}[1]{Eq.~\eqref{#1}}

\newcommand{\fref}[1]{Fig.~\ref{#1}}

\newcommand{\dd}{\mathrm{d}}
\newcommand{\fand}{\text{ and }}
\newcommand{\bs}{\boldsymbol}

\newcommand{\wt}{\widetilde}

\def\mnras{Mon. Not. R. Astron. Soc. }

\def\apj{Astrophys. J.}
\def\apjl{Astrophys. J. Lett.}
\def\prd{Phys. Rev. D}
\def\prl{Phys. Rev. Lett.}

\def\araa{Annu. Rev. Astron. Astrophys.}

\def\baas{Bull. Am. Astron. Soc.}

\begin{document}

\preprint{APS/123-QED}

\title{Phase shift of gravitational waves induced by aberration}

\author{Alejandro Torres-Orjuela}
\affiliation{Astronomy Department, School of Physics, Peking University, 100871 Beijing, China}
\affiliation{Kavli Institute for Astronomy and Astrophysics at Peking University, 100871 Beijing, China}

\author{Xian Chen}
\email{Corresponding author: xian.chen@pku.edu.cn}
\affiliation{Astronomy Department, School of Physics, Peking University, 100871 Beijing, China}
\affiliation{Kavli Institute for Astronomy and Astrophysics at Peking University, 100871 Beijing, China}

\author{Pau Amaro-Seoane}
\affiliation{Institute of Space Sciences (ICE, CSIC) \& Institut d'Estudis Espacials de Catalunya (IEEC) at Campus UAB, 08193 Barcelona, Spain}
\affiliation{Kavli Institute for Astronomy and Astrophysics at Peking University, 100871 Beijing, China}
\affiliation{Institute of Applied Mathematics, Academy of Mathematics and Systems Science, Chinese Academy of Sciences, Beijing 100190, China}
\affiliation{Zentrum f{\"u}r Astronomie und Astrophysik, TU Berlin, 10623 Berlin, Germany}

\date{\today}

\begin{abstract}
The velocity of a gravitational wave (GW) source provides crucial information about its formation and evolution processes. Previous studies considered the Doppler effect on the phase of GWs as a potential signature of a time-dependent velocity of the source.  However, the Doppler shift only accounts for the time component of the wave vector, and in principle motion also affects the spatial components.  The latter effect, known as ``aberration'' for light, is analyzed in this paper for GWs and applied to the waveform modeling of an accelerating source. We show that the additional aberrational phase shift could be detectable in two astrophysical scenarios, namely, a recoiling binary black hole (BBH) due to GW radiation and a BBH in a triple system. Our results suggest that adding the aberrational phase shift in the waveform templates could significantly enhance the detectability of moving sources.
\end{abstract}


\maketitle

\section{Introduction}\label{sec:int}

Gravitational waves (GWs) from merging binary
black holes (BBHs) and neutron stars have been detected by the LIGO and Virgo
detectors~\cite{ligo_2015,virgo_2012} with a signal-to-noise ratio (SNR)
between $10$ and $30$~\cite{GWTC1}. In the future a much higher SNR of
$\rho>100$ could be achieved by the next-generation ground-based detectors like
the Einstein Telescope and Cosmic Explorer~\cite{et_2010,cosmic_explorer_2019},
as well as by space-based observatories such as LISA~\cite{lisa_2017}. With a
high SNR, we will be able to probe the physical conditions of black holes (BHs)
and neutron stars to unprecedented accuracy.

To correctly retrieve the physical parameters of a GW source, an accurate
waveform is needed. For the early inspiral phase, the computation of
the waveform is usually performed in the center-of-mass (CoM) frame, thus
implicitly assuming that the source is at rest relative to the
observer~\cite{blanchet_2006,santamaria_ohme_2010,hannam_schmidt_2014,buonanno_damour_1999}.
The later merger and ringdown phases, can be described by Numerical Relativity (NR) waveforms or their surrogate models~\cite{SXS_2019,RIT_2019,GT_2016,blackman_field_2017,varma_field_2019}. These
methods allow the motion of the CoM during the merger and ringdown, but do not consider any motion prior to the merger. However, almost all
astrophysical objects are constantly moving, and hence the templates
that LIGO/Virgo use today should be considered as an approximation.
The approximation is acceptable when the SNR is low, but in the future when the
SNR is high, velocities could produce a detectable signature.

It is well known that motion induces, inter alia, a Doppler shift to a wave signal. However, if the velocity does not change, the Doppler shift is constant. For GWs, depending on redshift or blueshift, the resulting signal would be indistinguishable from that of a nonmoving source but with a greater mass and at a larger distance (the redshift case) or a smaller mass and at a shorter distance (blueshift case) (e.g.~\cite{chen_li_2017}). Therefore, it is difficult to prove that a GW source is moving unless the velocity varies with time, i.e., the source is accelerating~\cite{bonvin_caprini_2017}.

Recent studies focus on two astrophysical scenarios in which the acceleration could be high enough to produce a detectable Doppler-shift signature. The first scenario stems from the fact that GW radiation is often anisotropic and carries linear momentum~\cite{bonnor_rotenberg_1961,peres_1962}. Consequently, a BBH could be accelerated (kicked) during the last few cycles of the merger~\cite{fitchett_1983}. The kick velocity could reach thousands of ${\rm km\,s^{-1}}$ for special cases~\cite{pretorius_2005,herrmann_hinder_2007a,herrmann_hinder_2007b,herrmann_hinder_2007c,koppitz_2007,baker_centrella_2006,baker_boggs_2008,campanelli_lousto_2006,campanelli_lousto_2007a,campanelli_lousto_2007b,healy_herrmann_2009}, and in more ordinary conditions the velocity is on average $400\,{\rm
km\,s^{-1}}$ (e.g.~\cite{amaro-seoane_chen_2016}). It has been shown that LISA and the advanced LIGO detectors could detect kick velocities of around $500\,{\rm
km\,s^{-1}}$ based on Doppler shifts and as small as $200\,{\rm
km\,s^{-1}}$ by including higher modes~\cite{gerosa_moore_2016,calderon-bustillo_clark_2018,chamberlain_moore_2019}. The second scenario is motivated by the theoretical prediction that a population of merging BBHs may come from triple systems, with the third body being either a star~\cite{wen_2003,naoz_2016,meiron_kocsis_2017,arca-sedda_2020} or a supermassive BH in the center of a galaxy~\cite{antonini_perets_2012,mckernan_ford_2012,addison_gracia-linares_2019,bartos_kocsis_2017,stone_metzger_2017,tagawa_haiman_2019}. In this case, the acceleration is induced by the orbital motion of the BBH around the third body. If the BBH is inside the LISA band and can be tracked with a reasonable SNR for a duration of several months to several years, the acceleration is also detectable~\cite{meiron_kocsis_2017,inayoshi_tamanini_2017,tamanini_klein_2019,wong_baibhav_2019}. 

Although it is the most studied effect for moving GW sources, Doppler shift only accounts for the transformation of the time component of a wave vector, between Lorentz frames. However, wave vectors are four dimensional and a full treatment of the problem should include the transformation of the spatial components. For light waves, it is known that the transformation is the standard Lorentz transformation and the effect is aberration~\cite{jackson_2009}. For GWs, the aberration effect also exists and we recently showed that it affects the detected GW amplitude when the source is moving at a constant velocity~\cite{torres-orjuela_chen_2019}. In this paper, we study the aberration effect for accelerating sources and show that it induces an additional phase shift to the waveform which enhances the detectability of the motion of the source. Throughout this paper, unless otherwise indicated, we
use geometrical units in which the gravitational constant and the speed of light are equal to one (i.e., $G=c=1$).

\section{The effect of motion on the phase}\label{sec:eff}

In order to calculate the effect of motion on the observed phase of GWs, we first specify the coordinate systems (COs) for the source and the observer. We define $S := (\hat{u},\hat{a},\hat{b},\hat{L})$ as the CO of the source. We write $\hat{u}$ for the 4-velocity of the source, which represents the direction of the time coordinate in the rest frame of the source. The other 4-vectors, $\hat{a}$, $\hat{b}$, and $\hat{L}$, are unit vectors representing the spatial coordinates, and hence set to be perpendicular to each other and to $\hat{u}$. We define the CO of the observer as $O := (\hat{v},\hat{d}_1,\hat{d}_2,\hat{p})$. Here $\hat{v}$ is the 4-velocity of the observer, representing the time coordinate of $O$. Moreover, $\hat{d}_1$, $\hat{d}_2$, and $\hat{p}$ are three unit 4-vectors representing the spatial coordinates in the rest frame of the observer. Therefore, they are set to be perpendicular to each other and to $\hat{v}$.

Without losing generality, we set the spatial coordinates $\hat{a}$ and $\hat{b}$ in $S$ to be in the
plane perpendicular to the angular-momentum vector of the GWs source and $\hat{L}$ to be aligned with the angular momentum. Further, we set $\hat{d}_1$, $\hat{d}_2$ and $\hat{p}$ to be aligned with $\hat{a}$, $\hat{b}$ and $\hat{L}$, respectively, in the limit of a vanishing relative velocity
between $S$ and $O$. In this way, 4-vectors transform between the two COs as
\begin{equation}\label{eq:cor}
S \ \genfrac{}{}{0pt}{1}{\underrightarrow{\ \Lambda(-\bs{u}) \ }}{\overleftarrow{\ \ \Lambda(\bs{u}) \ \ }} \ O,
\end{equation}
where $\Lambda$ denotes the Lorentz transformation and $\bs{u}$ is the 3-velocity of the source seen in $O$. Unless otherwise indicated, for any 4-vector (e.g., $\hat{x}$), we denote the related 3-vector, relative to the specified rest frame, by the same symbol written in bold letters (e.g., $\bs{x}$).

Having specified the COs we define the wave vector, $\hat{k}$, respective to $S$ and $O$ as
\begin{equation}\label{eq:wvc}
\hat{k}_S := \omega\left(\begin{array}{c} 1 \\ \bs{n} \end{array}\right) \fand \hat{k}_O := \wt{\omega}\left(\begin{array}{c} 1 \\ \bs{m} \end{array}\right),
\end{equation}
where $\bs{n}$ and $\bs{m}$ are 3-vectors of unit length, representing the line of sight (LoS) from the source to the observer in the respective COs. Moreover, $\omega$ and $\wt{\omega}$ are the frequencies of the GW in $S$ and $O$. 

Lorentz transformation changes the time component of a wave vector, according to the standard Doppler shift, as
\begin{equation}\label{eq:ftf}
\wt{\omega} = \frac{\omega}{\gamma(1-\langle\bs{m},\bs{u}\rangle)},
\end{equation}
where $\langle\cdot,\cdot\rangle$ denotes the 3-dimensional Euclidean scalar product and $\gamma := (1-u^2)^{-1/2}$ is the Lorentz factor to the velocity of magnitude $u := \sqrt{\langle\bs{u},\bs{u}\rangle}$. The shift of frequency leads to a shift in the GW phase as
\begin{equation}\label{eq:dps}
\Phi_\text{Dop}(t) = \int^{t}_{t_s} \frac{\omega(t')}{\gamma(t')(1-\langle\bs{m},\bs{u}(t')\rangle)}\dd t',
\end{equation}
where $t_s$ denotes the time in $O$ at which the observation starts. The above equation allows $\bs{u}$ to be time-dependent, and it has been used to calculate the phase shift of an accelerating GW source.  In principle, the LoS $\bs{m}$ could be time-dependent, too. However, given the large distances of
GW sources~\cite{GWTC1}, it is unlikely to see a change of the source's sky location during its observation. Therefore, we set $\bs{m}$ to be constant.

\begin{figure}[tpb] \centering \includegraphics[width=0.48\textwidth]{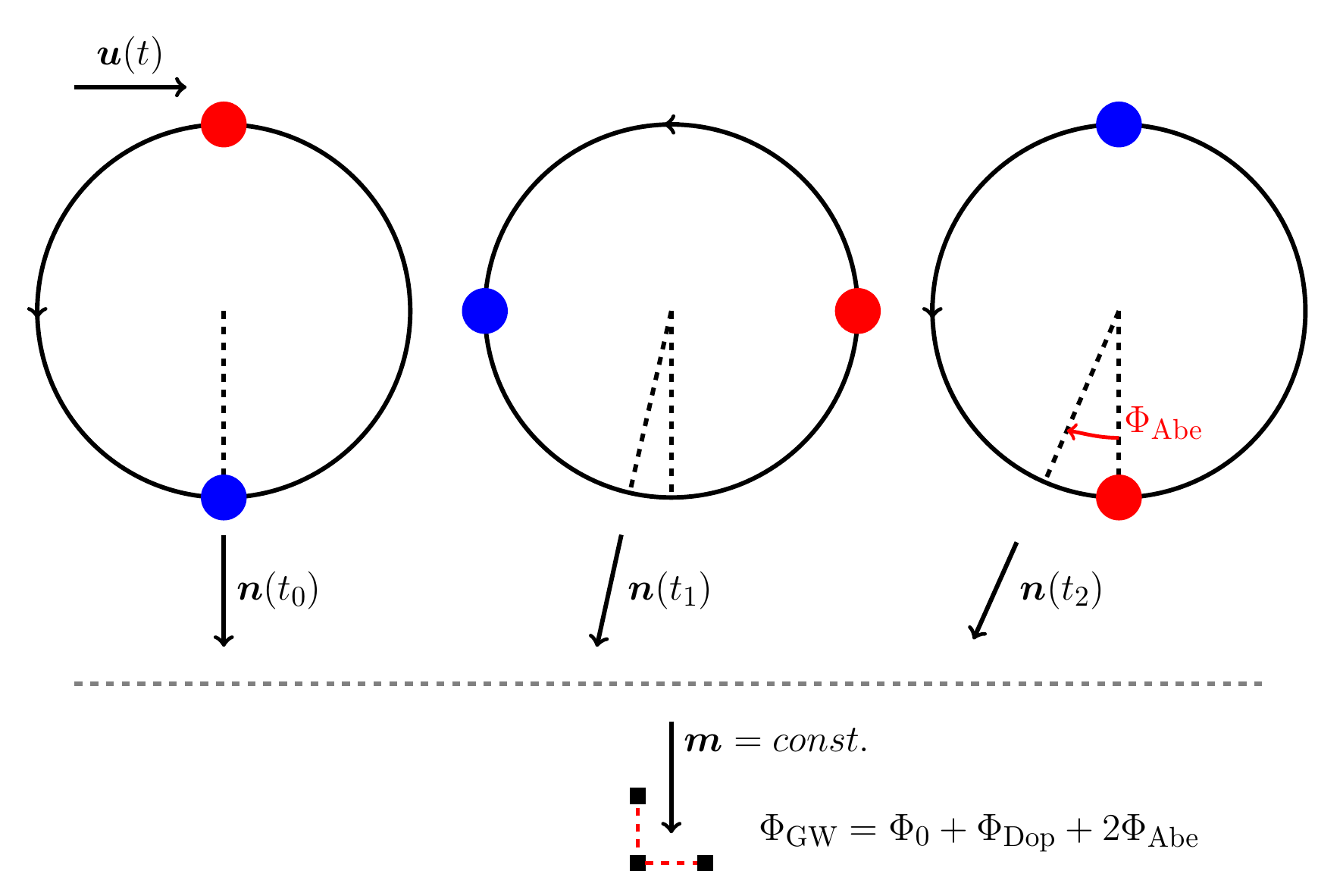}
\caption{
Aberration of GWs, induced by an increasing velocity of the source, and its effect on the phase detected by a distant observer. For the distant observer the LoS $\bf{m}$ is fixed, but in the source frame the corresponding LoS $\bf{n}$ is changing because of the acceleration. This change of the LoS in the source frame leads to a shift of the GW phase detected by the distant observer.
}\label{fig:aps}
\end{figure}

Lorentz transformation also changes the spatial components of a wave vector. For constant velocity, the spatial components do not change with time. However, when the velocity varies, the LoS $\bs{n}$ in $S$ changes even if the LoS $\bs{m}$ in $O$ does not change. In \fref{fig:aps} we show how this change of $\bs{n}$ affects the GW phase using the example of a BBH. The variation of the LoS $\bs{n}$ induces a shift to the apparent phase of the orbit and consequently two times the same shift to the apparent phase of the GW for the dominant $(2,2)$ mode we consider here.

To calculate the orbital phase shift due to the aberration effect, $\Phi_\text{Abe}$, we use that $\hat{k}_S$ and $\hat{k}_O$ in \eq{eq:wvc} are related by Lorentz transformations. We find that the LoS in $S$ is
\begin{equation}\label{eq:trwv}
\bs{n}(t) = \frac{\bs{m} - \gamma(t)\bs{u}(t) + (\gamma(t)-1)\langle\bs{m},\bs{u}(t)\rangle\bs{u}(t)/u(t)^2}{\gamma(t)(1-\langle\bs{m},\bs{u}(t)\rangle)},
\end{equation}
and for the normalized projection of $\bs{n}$ in the orbital plane of the binary
\begin{equation}\label{eq:dpv}
\bs{q}(t) := \frac{1}{\sqrt{n_a(t)^2 + n_b(t)^2}}\left(\begin{array}{c} n_a(t) \\ n_b(t) \\ 0 \end{array}\right),
\end{equation}
where $n_a$ and $n_b$ are the projections of $\bs{n}$ along $\bs{a}$ and $\bs{b}$ in $S$. Then, given $\bs{q}_s$ as $\bs{q}$ at the time the observation of the GW signal starts, we can calculate the
aberrational phase shift with
\begin{equation}\label{eq:aps}
\Phi_\text{Abe}(t) = \arccos{(\langle\bs{q}_s,\bs{q}(t)\rangle)}.
\end{equation}

Combining the phase shifts due to the Doppler and the aberration effects, we have
\begin{equation}\label{eq:dgwp}
\Phi_\text{GW}(t) = \Phi_0 + \Phi_\text{Dop}(t) \pm 2\Phi_\text{Abe}(t),
\end{equation}
where $\Phi_0$ is a constant phase shift. The sign before the term $2\Phi_\text{Abe}$ is plus/minus if the rotation of $\bs{q}$ is retrograde/prograde with respect to the orbital motion of the
binary.

We remark that the direct cause of the aberrational phase shift differs from the cause of the Doppler shift. The former is induced by the change of the LoS in $S$, in which the GW rays must point to reach
the observer. The Doppler phase shift is related to the frequency of the GWs and thus is caused by a different flow of time in the two systems, $S$ and $O$.

\section{Comparing the phase shifts}\label{sec:com}

In most astrophysical scenarios, the velocity of the GW source is small compared to the speed of light. Therefore, in this section we assume $u \ll 1$ and consider the Doppler and aberrational phase shifts in this limit.

Expanding the Doppler phase shift $\Phi_\text{Dop}$ to first order in $u$, we find
\begin{equation}\label{eq:dpse}
\Phi_\text{Dop}(t) = \int^t_{t_s} (1 + \langle\bs{m},\bs{u}(t')\rangle)\omega(t')\dd t'.
\end{equation}
We see that the Doppler phase shift is, to the lowest order, proportional to $u$ and, more importantly, only proportional to the component of the velocity along the LoS, as is indicated by the term $\langle\bs{m},\bs{u}(t')\rangle$.

Next, we also expand the aberrational phase shift $\Phi_\text{Abe}$ to first order in $u$. We use that $\arccos{(1-x)} \approx \sqrt{2x}$, for $0 < x \ll 1$, and expand accordingly $\bs{q}_s$ and $\bs{q}$ to second order in $u$. We find that
\begin{equation}\label{eq:apse}
\Phi_\text{Abe}(t) = \mu\sqrt{\langle\tilde{u}(t),\tilde{u}(t)\rangle - \mu^2\langle\wt{m},\tilde{u}(t)\rangle^2},
\end{equation}
where $\tilde{m} := (m_1,m_2,0)^\text{T}$ is the projection of the LoS in the plane defined by $\bs{d}_1$ and $\bs{d}_2$ in $S$, $\mu := \sqrt{\langle\tilde{m},\tilde{m}\rangle}^{-1}$ is the inverse of the length of $\tilde{m}$, and $\tilde{u}(t) := (u_1(t)-u_1(t_s),u_2(t)-u_2(t_s),0)^\text{T}$ calculates the change of the velocity of the source from the time when the observation starts, $t_s$, to the actual time, $t$, and  projects it into the $\bs{d}_1-\bs{d}_2$ plane. We note that in order to expand $\Phi_\text{Abe}$ we have assumed $\mu \lesssim 10$. This condition can be fulfilled in most cases because it requires that the inclination of the orbital plane of the source is $\gtrsim 5^\circ$, where $0^\circ$ corresponds to a face-on configuration. Now it is clear from \eq{eq:apse} that the aberrational phase shift is to the lowest order proportional to $u$. It vanishes only when the projected velocity difference, $\tilde{u}$, is parallel to the projected LoS, $\tilde{m}$, and it becomes maximal when the two projected vectors are perpendicular.

We find that the aberrational and the Doppler phase shifts are both proportional to the magnitude of the velocity of the source. Therefore, they are equally important to our understanding of the effect of motion on the GW signal.

As a next step, we calculate the phase shift of the GWs in the aforementioned
two astrophysical scenarios, i.e, (i) a merging BBH receiving a gravitational
kick and (ii) a BBH orbiting a third body. Previous studies have
considered the Doppler effect and showed that the phase shift is
detectable~\cite{gerosa_moore_2016,inayoshi_tamanini_2017,meiron_kocsis_2017}.
As for the aberrational phase shift, in the above scenario (i) it is
automatically included in models containing higher modes and fitted to NR
waveforms~\cite{varma_field_2019,cotesta_buonanno_2018}. Nevertheless, our
method is different from NR and provides an alternative and computationally
inexpensive way of including the aberrational phase shift into the waveforms of
merging BBHs.  For scenario (ii), however, the aberrational phase shift has not
been considered in previous works. In the following, we will show that
including the aberration effect in the waveform makes the detection of the
motion even easier.

\begin{figure}[tpb] \centering \includegraphics[width=0.48\textwidth]{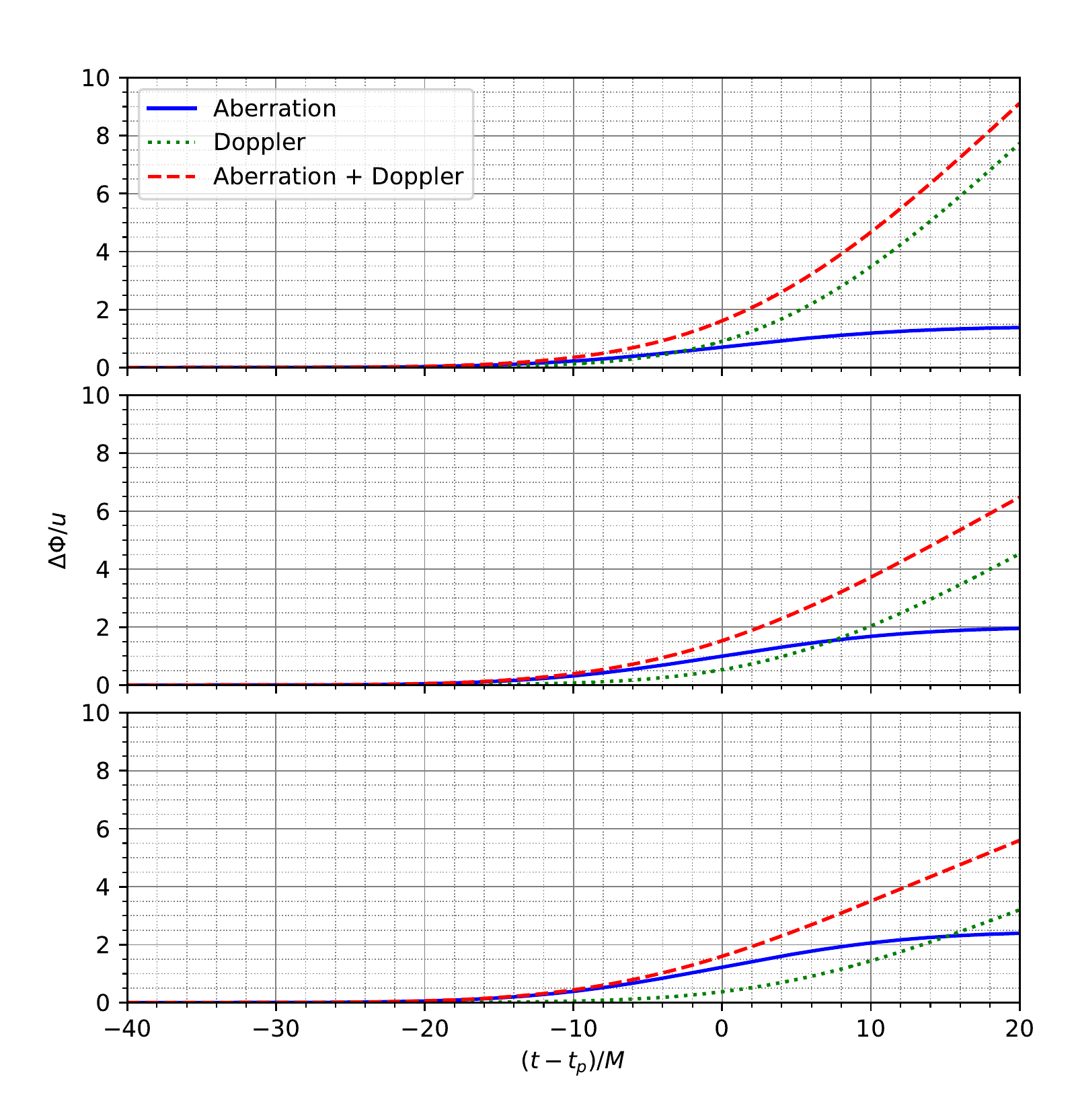}
\caption{
Evolution of the aberrational, Doppler and total phase shifts of a kicked source. The results are normalized by the final	kick velocity. The top, middle and bottom panel correspond to angles of $30^\circ$, $60^\circ$, and $70^\circ$ between the kick	velocity and the LoS in $O$.  
}\label{fig:pdk} 
\end{figure}

For the kick, we adopt a model where the direction of the velocity is fixed and the magnitude builds up following a Gaussian distribution~\cite{gerosa_moore_2016}. The Gauss distribution is set to be centered at the time of the peak GW amplitude and to have a variance of $10~M$~\cite{brugmann_gonzalez_2008,luosto_zlochower_2008}, where $M$ is the total mass of the system.
Moreover, we adopt the model presented in Ref.~\cite{mcwilliams_2019} for an equal-mass binary
to compute the evolution of the GW frequency in $S$. The results are shown in \fref{fig:pdk}, where the phase shifts are normalized by the final kick velocity, to make the result scalable to other kick velocities, and the time is normalized by $M$. The three different panels correspond to different angles between the velocity and the LoS in $O$. In all these cases the aberrational phase shift is comparable to the Doppler phase shift. In the two cases with larger angles, the aberrational phase shift even predominates at times $t \lesssim 10~M$. Most importantly, the total phase shift, including the aberrational one, is significantly greater than the Doppler phase shift by a factor of about 1.2, 1.5, and 1.7, respectively, in the three cases.

\begin{figure}[tpb] \centering \includegraphics[width=0.48\textwidth]{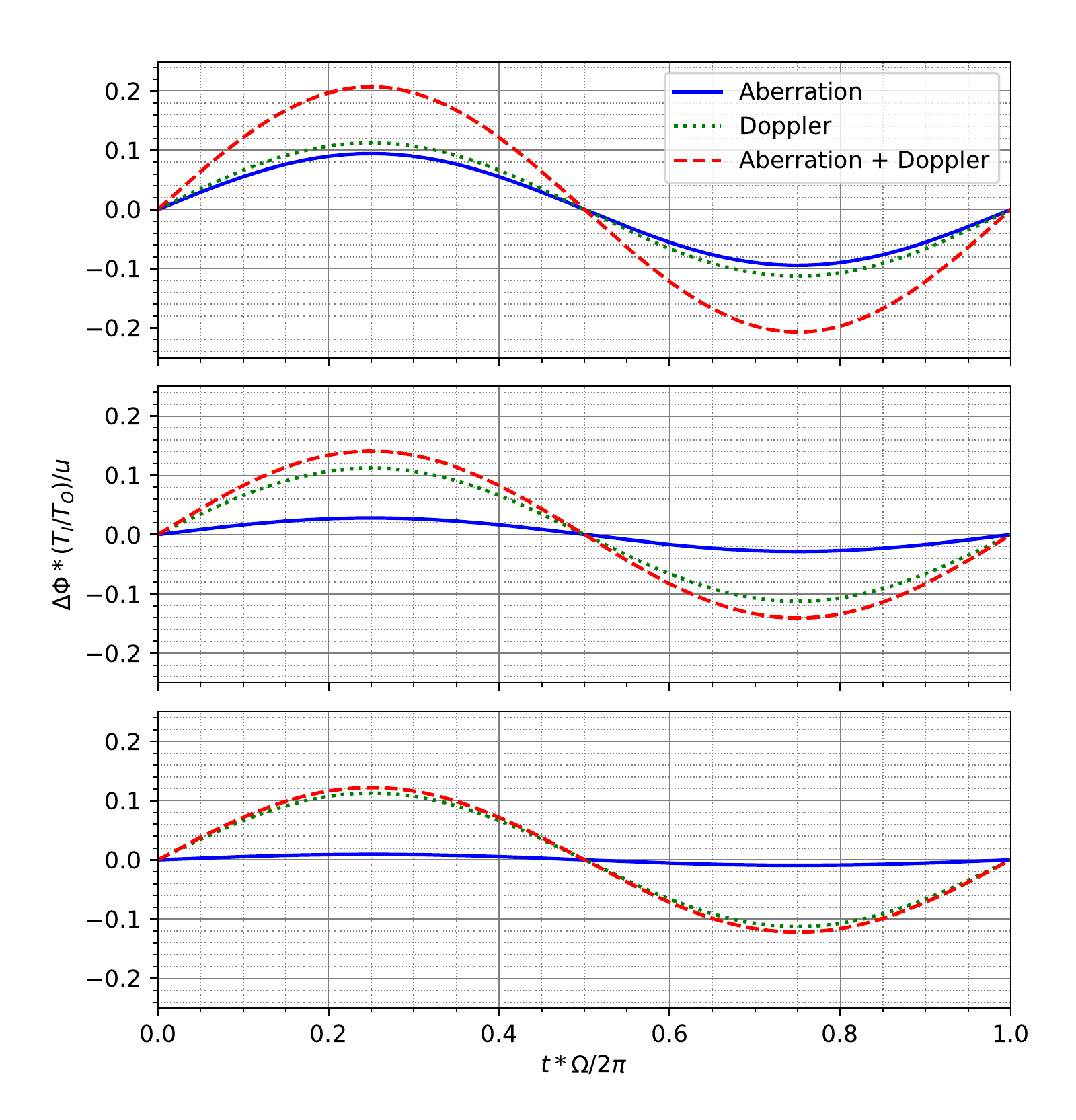}
\caption{
Aberrational, Doppler and total phase shift of a BBH orbiting a third body. The phase shifts are normalized, using the ratio of periods of the outer and inner orbits ($T_O/T_I$) as well as the magnitude of the outer orbital velocity ($u$). The top,	middle and bottom panel correspond to, respectively, $T_O/T_I=30$,	$100$ and $300$. The time axis is normalized to show the phase of the outer orbit.
}\label{fig:pdo}
\end{figure}

For a BBH orbiting a third body, we assume for simplicity that both the inner (the BBH) and the outer (the binary formed by the BBH and the third) binaries have circular orbits and constant orbital periods. Moreover, the two orbital planes are coplanar in $S$ (see Ref.~\cite{meiron_kocsis_2017} for a similar configuration). \fref{fig:pdo} shows the shift of the GW phase of the inner BBH over one period around the third body. The results are normalized to highlight the dependence on the velocity of the outer orbit as well as the ratio of the periods of the inner and outer binaries. In the three panels, the outer orbits have the same inclination of $45^\circ$ in $O$, while the ratio of the outer and inner orbits is $30$, $100$, and $300$ from the top to the bottom panel. In the top panel, the aberrational and the Doppler phase shift have a similar magnitude, thus contributing almost equally to the total phase shift. As the orbital periods of the inner and outer binary become more different, we see in the lower two panels that the contribution of the aberration effect to the total phase shift diminishes. This is because the Doppler shift is proportional to the instantaneous speed, while the aberrational shift is proportional to its change in time.

\section{Detectability}\label{sec:det}

Whether the phase shifts shown above are detectable depends on the loudness of the signal, which we characterize by its SNR. In the following, we adopt the scheme presented in Ref.~\cite{lindblom_owen_2008} to derive the SNR which is needed to distinguish an accelerating GW source from a stationary one.

Suppose the exact signal of a source $h_e$ differs from a model signal by a logarithmic amplitude $\delta\chi$ and a phase shift $\delta\Phi$. The two signals are distinguishable if
\begin{equation}\label{eq:sdc}
\overline{\delta\chi}^2 + \overline{\delta\Phi}^2 > {1}/{\rho^2},
\end{equation}
where $\rho^2 := \langle h_e|h_e\rangle$ is the SNR of the exact signal, $\overline{\delta\chi}^2 :=
\langle\delta\chi\hat{h}_e|\delta\chi\hat{h}_e\rangle$ and $\overline{\delta\Phi}^2 :=
\langle\delta\Phi\hat{h}_e|\delta\Phi\hat{h}_e\rangle$ are the signal-weighted averages of the logarithmic amplitude and phase errors, $\hat{h}_e := h_e\rho^{-1}$ is the normalized exact signal, and $\langle\cdot|\cdot\rangle$ denotes the noise-weighted inner product. For our problem, we can neglect the difference in the amplitude because for nonrelativistic velocities the difference is
small~\cite{torres-orjuela_chen_2019,lindblom_owen_2008}. Moreover, since a real GW signal has a finite interval of frequencies, we can use the first mean value theorem for definite integrals~\cite{comenetz_2002} to extract the phase difference out of the noise-weighted inner product. These considerations and the fact that $\langle\hat{h}_e|\hat{h}_e\rangle = 1$ help us to simplify the above criterion to
\begin{equation}\label{eq:ddc}
{\delta\Phi(\omega_r)}>1/\rho,
\end{equation}
where $\omega_r$ is a representative value within the frequency range of the signal. To get $\delta\Phi(\omega_r)$, we Fourier transform the phase-shifted waveforms corresponding to \fref{fig:pdk} and \fref{fig:pdo} and use the mean value of all frequencies as an approximation to $\omega_r$.

\begin{figure}[tpb] \centering \includegraphics[width=0.48\textwidth]{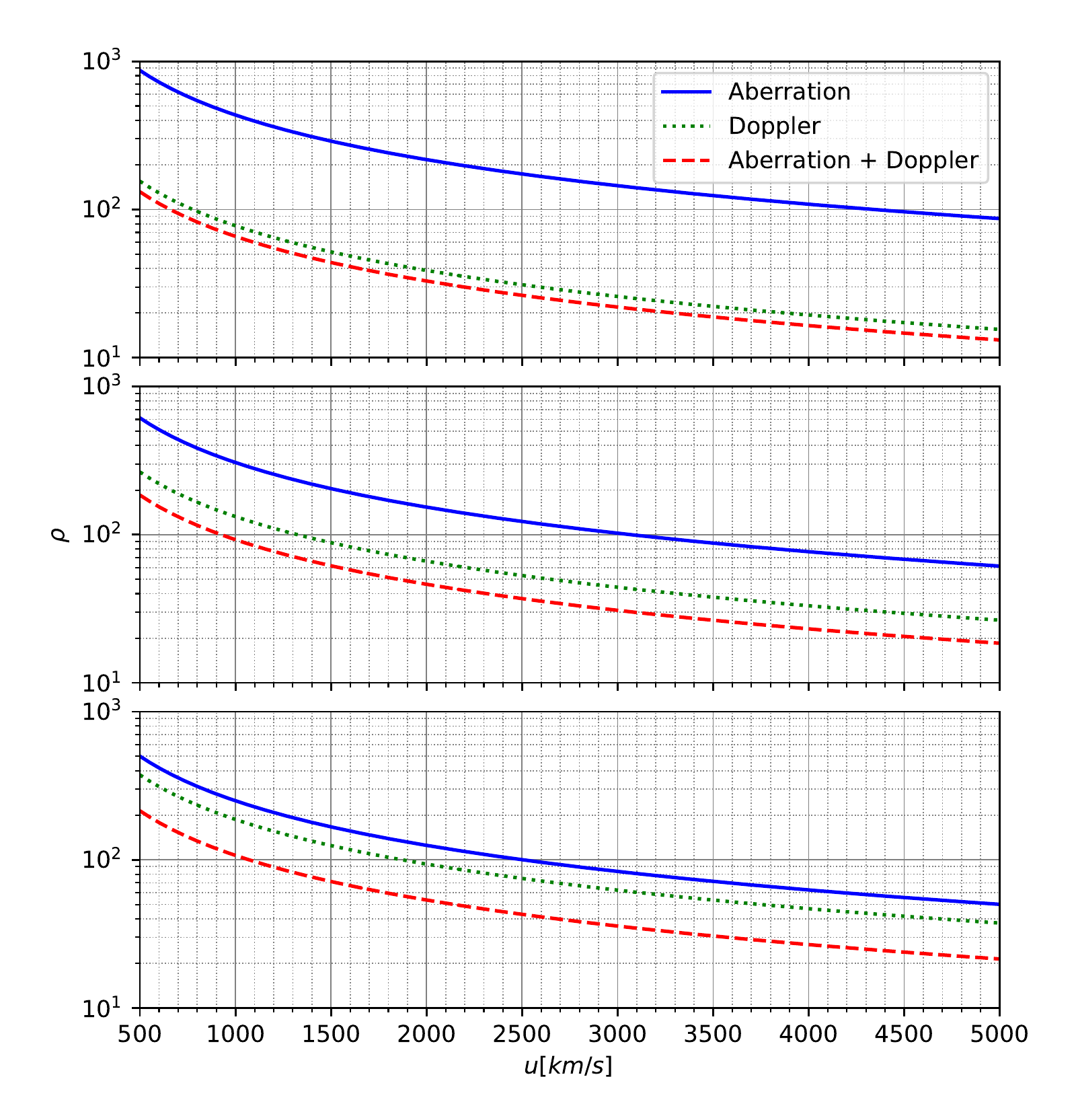}
\caption{
SNR required to detect the aberrational, Doppler and total phase shifts 
	of a kicked source. The three subplots correspond to the three orientations of the velocity relative to the LoS shown in \fref{fig:pdk}.}
\label{fig:snrk}
\end{figure}

\fref{fig:snrk} shows the SNR required to detect the aberrational, Doppler and total phase shifts. The three panels correspond to the three cases presented in \fref{fig:pdk}. The SNR, as is expected, is a decreasing function of the final kick velocity $u$. When both the aberrational and the Doppler effects are included, the SNR needed to detect the phase shift is significantly lower than the SNR accounting for only the Doppler effect, by a factor of about $1.2$, $1.5$, and $1.7$ in the three cases from top to bottom. The lowering of the required SNR has two important astrophysical implications. (i) Given an SNR, it allows the detection of a significantly smaller kick velocity. Meanwhile, ignoring the aberrational phase shift could lead to an overestimation of the kick velocity, in order to explain the larger phase shift. (ii) It increases the horizon distance of a particular source. The number of detectable sources scales with the detection volume and hence increases even faster, by a factor of $1.7$, $3.4$, and $4.9$, correspondingly.

\begin{figure}[tpb] \centering \includegraphics[width=0.48\textwidth]{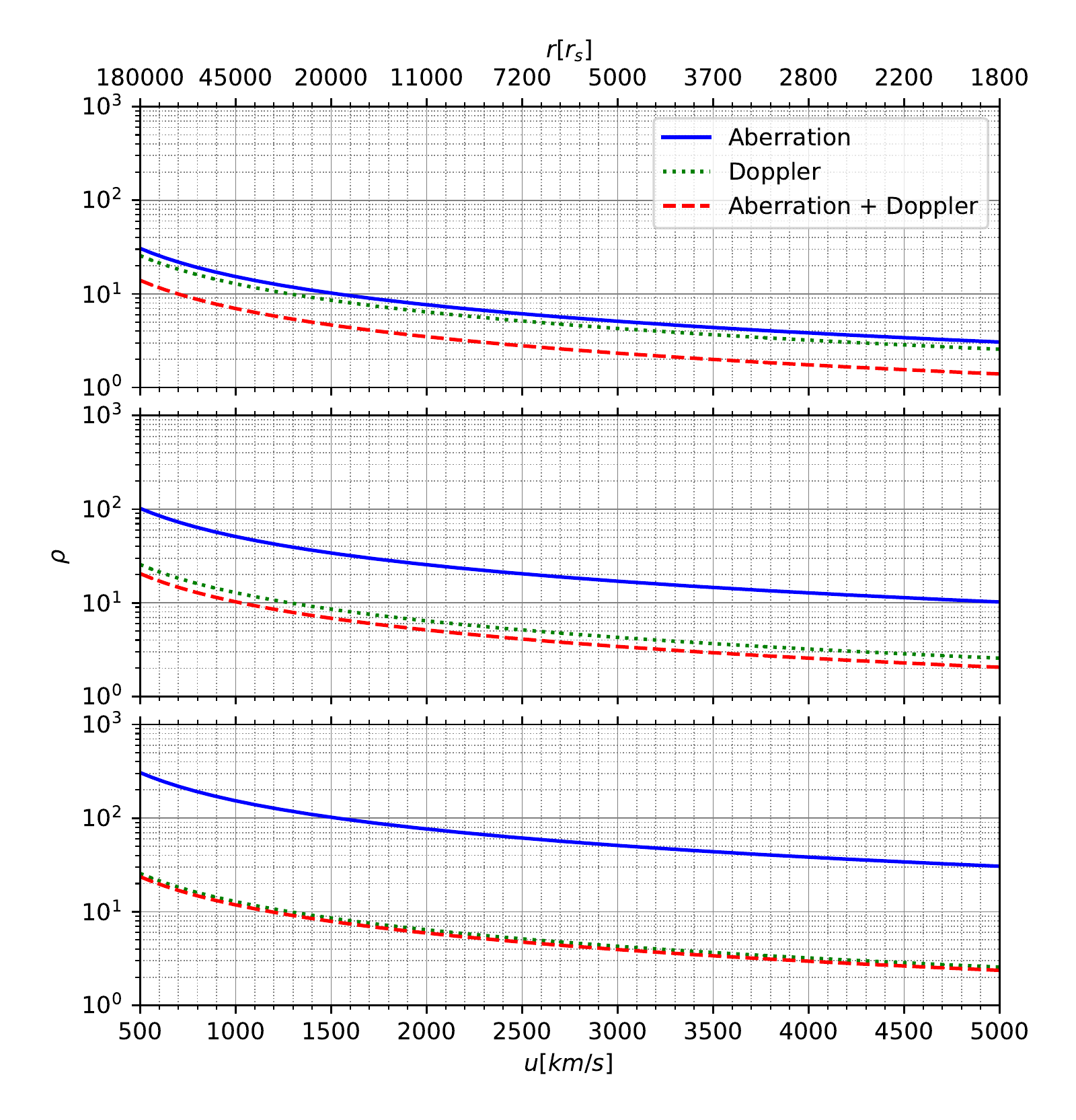}
\caption{
SNR required to extract the aberrational, Doppler and total phase shifts 
	of a BBH orbiting a third body. The upper abscissa shows the corresponding distance to the third body in units of its Schwarzschild radius. The three subplots correspond to the three ratios of the periods shown in \fref{fig:pdo}.
}\label{fig:snro}
\end{figure}

\fref{fig:snro} shows the required SNR corresponding to the three cases presented in \fref{fig:pdo}, where we have assumed that the inner binary emits GWs at a frequency of $1$ mHz and the observational period is three years (corresponding to LISA). Again we find that the SNR required to detected the
total phase shift is lowered compared to the SNR required to detect only the Doppler phase shift, by a factor of $1.8$, $1.3$, and $1.1$ in the three panels from top to bottom. This less demanding SNR would allow the detection of a smaller velocity, by the same factor, and increase the number of detectable sources by a factor of about 5.8, 2.2 and 1.3, respectively.

We show that considering the aberrational phase shift in addition to
the Doppler phase shift allows an improvement in the detectability of
velocities comparable to the one achieved by considering higher modes for
kicks~\cite{calderon-bustillo_clark_2018}. This improvement comes for
relatively low computational costs and can further be applied to the case of a
motion induced by a third body, for which there exist no models including
higher modes up to date.

\section{Conclusions}\label{sec:con}

We find that the aberrational phase shift,
induced by a time-dependent velocity of the GW source, is of the same order as
the Doppler shift and could significantly enhance the detectability of
the motion of a GW source. Considering the total phase shift of the dominant
(2,2) mode provides a computationally inexpensive method of detecting the
motion. Moreover, the method could be extended to higher modes to further
improve the accuracy of waveforms modeling the motion. Ignoring the
aberrational phase shift, on the other hand, could lead to an overestimation
of the velocity of the source along the LoS and, maybe more seriously,
misinterpret the physical or astrophysical origin of such a velocity.

\section*{Acknowledgments}

This work is supported by the National
Science Foundation of China grants No~11873022 and 11991053. A.T.O. is partly
supported by the Strategic Priority Research Program of the Chinese Academy of
Sciences, Grant No.  XDB23040100 and No. XDB23010200. P.A.S. acknowledges
support from the Ram{\'o}n y Cajal Programme of the Ministry of Economy,
Industry and Competitiveness of Spain, as well as the COST Action GWverse
CA16104.

\bibliographystyle{apsrev4-1.bst}

\end{document}